\documentclass[aps,prc,reprint,showpacs,groupedaddress]{revtex4-1}
\usepackage{amssymb}
\usepackage{graphicx}
\usepackage{color}

\begin{document}

\newcommand{\be}{\begin{equation}}
\newcommand{\ee}{\end{equation}}
\newcommand{\ba}{\begin{eqnarray}}
\newcommand{\ea}{\end{eqnarray}}
\newcommand{\lp}{\left(}
\newcommand{\rp}{\right)}
\newcommand{\ApJ}{Astrophys. J.}
\newcommand{\MNRAS}{Mon. Not. R. Astron. Soc.}
\newcommand{\AandA}{Astron. Astrophys.}
\newcommand{\kB}{k_{\mathrm{B}}}
\newcommand{\kappagl}{\kappa}
\newcommand{\mpr}{m_{\mathrm{p}}}
\newcommand{\mn}{m_{\mathrm{n}}}
\newcommand{\mpeff}{\mpr^{\ast}}
\newcommand{\mneff}{\mn^{\ast}}
\newcommand{\nb}{n_{\mathrm{b}}}
\newcommand{\nel}{n_{\mathrm{e}}}
\newcommand{\npr}{n_{\mathrm{p}}}
\newcommand{\nn}{n_{\mathrm{n}}}
\newcommand{\nnuc}{n_{\mathrm{nuc}}}
\newcommand{\Tcp}{T_{\mathrm{cp}}}
\newcommand{\kf}{k_{\mathrm{F}}}
\newcommand{\ef}{\varepsilon_{\mathrm{F}}}
\newcommand{\hczero}{H_{\mathrm{c}}}
\newcommand{\hcone}{H_{\mathrm{c1}}}
\newcommand{\hctwo}{H_{\mathrm{c2}}}
\newcommand{\phizero}{\phi_0}
\newcommand{\tohm}{\tau_{\mathrm{OhmH}}}
\newcommand{\tohmzero}{\tau_{\mathrm{Ohm}}}
\newcommand{\tsc}{\tau_{\mathrm{sc}}}
\newcommand{\econd}{\sigma_{\mathrm{c}}}
\newcommand{\lmag}{l_{\mathrm{mag}}}
\newcommand{\tcool}{\tau_{\mathrm{cool}}}
\newcommand{\heatcap}{C}
\newcommand{\Msun}{M_{\mathrm{Sun}}}
\newcommand{\Mdu}{M_{\mathrm{dU}}}
\newcommand{\lcool}{l_{\mathrm{cool}}}


\title{Dynamical onset of superconductivity and \\ retention of magnetic fields in cooling neutron stars}
\author{Wynn C.~G. Ho$^{1,2}$,
Nils Andersson$^1$,
and Vanessa Graber$^3$}
\affiliation{$^1$Mathematical Sciences and STAG Research Centre, University of Southampton, Southampton, SO17 1BJ, United Kingdom \\
$^2$Physics and Astronomy, University of Southampton, Southampton, SO17 1BJ, United Kingdom \\
$^3$Department of Physics and McGill Space Institute, McGill University, Montreal, QC, H3A 2T8, Canada}

\date{Received \today; revised manuscript received ; published}

\begin{abstract}
A superconductor of paired protons is thought to form in the core of neutron
stars soon after their birth.
Minimum energy conditions suggest magnetic flux is expelled from the
superconducting region due to the Meissner effect, such that the neutron
star core is largely devoid of magnetic fields
for some nuclear equation of state and proton pairing models.
We show via neutron star cooling simulations that the superconducting region
expands faster than flux is expected to be expelled
because cooling timescales are much shorter than timescales of magnetic
field diffusion.
Thus magnetic fields remain in the bulk of the neutron star core for at least
$10^6-10^7\mbox{ yr}$.
We estimate the size of flux free regions at $10^{7}\mbox{ yr}$ to be
$\lesssim 100\mbox{ m}$ for a magnetic field of $10^{11}\mbox{ G}$ and
possibly smaller for stronger field strengths.
For proton pairing models that are narrow, magnetic flux may be completely
expelled from a thin shell of approximately the above size after
$10^5\mbox{ yr}$.
This shell may insulate lower conductivity outer layers, where magnetic
fields can diffuse and decay faster, from fields maintained in the highly
conducting deep core.
\end{abstract}


\maketitle

\section{Introduction \label{sec:intro}}

Neutron stars (NSs) are unique probes of the dense matter equation of state
(EOS), which prescribes a relationship between pressure and density and
determines the behavior of matter near and above nuclear densities
($\nnuc\approx 0.16\mbox{ fm$^{-3}$}$).
For example, some EOSs predict the presence of exotic particles, such as
hyperons and deconfined quarks, in the NS inner core
at baryon number densities $\nb>\nnuc$
(see, e.g., \cite{haenseletal07,lattimerprakash16}, for review).
At the same time, theory and observations indicate that the core of NSs
(at $\nb\gtrsim 0.1\mbox{ fm$^{-3}$}$) may contain a neutron superfluid and
proton superconductor
\cite{pageetal11,shterninetal11,elshamoutyetal13,posseltetal13}.

In this present work, we are concerned with the onset of proton
superconductivity, which takes place when the local temperature $T$ falls
below the proton critical temperature $\Tcp$
(see, e.g., \cite{sauls89,pageetal14}, for review).
The latter is related to the energy gap for Cooper pairing $\Delta$
in the zero temperature limit
by $\kB\Tcp\approx 0.5669\Delta$ for singlet ($^1S_0$) pairing.
A paired proton superconductor can take two forms in the core of a NS,
depending on the Ginzburg-Landau parameter
\be
\kappagl \equiv \lambda/\xi \label{eq:ginzburglandau}
\ee
and two critical magnetic fields
\be
\hcone = \frac{\phizero}{4\pi\lambda^2}\ln\kappagl
\qquad \mbox{and}\qquad \hctwo=\frac{\phizero}{2\pi\xi^2},
\label{eq:hc1}
\ee
where $\phizero=\pi\hbar c/e$ is the magnetic flux quantum
and the equation for $\hcone$ is in the limit of large $\kappagl$
\cite{tinkham96}.  The magnetic field penetration lengthscale is
\be
\lambda =\lp\frac{\mpeff c^2}{4\pi e^2\nel}\rp^{1/2} \label{eq:lmag}
\ee
and the superconductor pairing or coherence lengthscale (also typical size
of magnetic fluxtube) is
\be
\xi = \frac{2\ef}{\pi\kf\Delta} = \frac{\hbar^2\kf}{\pi\mpeff\Delta},
 \label{eq:lsc}
\ee
where $\mpeff$ is effective proton mass,
$\ef$ and $\hbar\kf=\hbar(3\pi^2\npr)^{1/3}$ are Fermi energy and momentum,
respectively, and $\nel$ and $\npr$ are electron and proton number densities,
respectively.

If the Ginzburg-Landau parameter $\kappagl<1/\sqrt{2}$, then an external
magnetic field $H$ does not penetrate significantly into the superconductor,
and magnetic flux is expelled from superconducting regions (such that $B=0$)
due to the Meissner effect
(see, e.g., \cite{baymetal69,sauls89,tinkham96,graberetal17}).
In this state, magnetic flux can be retained in macroscopic regions of normal
conducting matter that alternate with regions of flux-free superconducting
matter.
Conversely, if $\kappagl>1/\sqrt{2}$, then magnetic field can reside in
superconducting fluxtubes.
Substituting Eqs.~(\ref{eq:lmag}) and (\ref{eq:lsc}) into
Eq.~(\ref{eq:ginzburglandau}), we find
\be
\kappagl \approx 0.8\lp\Delta/\mbox{1 MeV}\rp\lp\npr/\nnuc\rp^{-5/6}.
\ee
The energy of the fluxtube state is at a minimum when the magnetic field
$H$ is $\hcone\lesssim H\lesssim\hctwo$.
For $H\lesssim\hcone$, the superconductor should be in a Meissner state
(i.e., magnetic flux expulsion),
while superconductivity is destroyed for $H\gtrsim\hctwo$.

\begin{figure}
\resizebox{1.0\hsize}{!}{\includegraphics{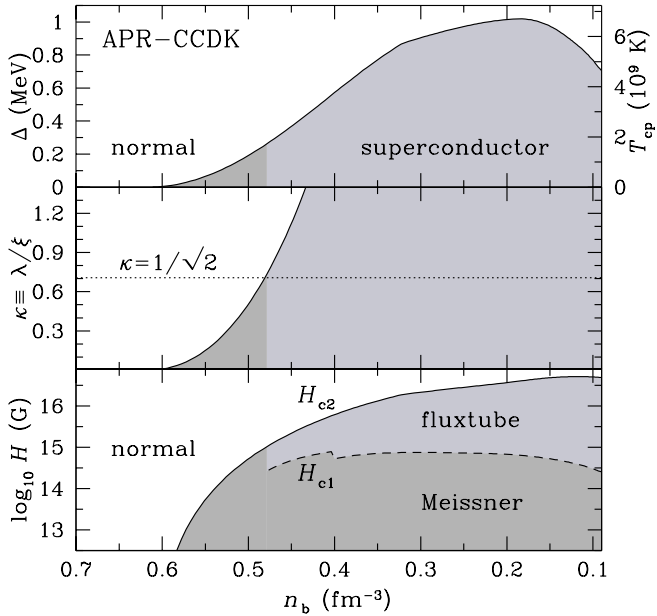}}
\caption{
Proton superconductor states.
Top: CCDK model of singlet pairing energy gap $\Delta$ (left axis)
and critical temperature $\Tcp$ (right axis) as a function
of baryon number density $\nb$, calculated using the APR EOS model.
Middle: Ratio $\kappagl$ between magnetic field penetration lengthscale
$\lambda$ and fluxtube size $\xi$.
The two shaded regions separated at $\nb=0.48\mbox{ fm$^{-3}$}$ denote the
regime where only the Meissner (flux expulsion) state is allowed
(when $\kappagl<1/\sqrt{2}$)
and the regime where the superconductor can be in Meissner state or in fluxtubes
(when $\kappagl>1/\sqrt{2}$).
Bottom: Dependence of superconductor state on $\nb$ and magnetic field
$H$.  Superconductivity is destroyed when $H>\hctwo$.
The fluxtube state exists when $\kappagl>1/\sqrt{2}$ and
$\hcone\lesssim H\lesssim\hctwo$, and the Meissner state exists otherwise.
The behavior of $\hcone$ for $\kappagl<2$ is approximated using results from
\cite{tinkham96}.
\label{fig:gap_apr_ccdk}}
\end{figure}

These conditions on $\kappagl$ and $H$ (relative to $\hcone$ and $\hctwo$)
determine in which regions in a NS are superconducting protons in a fluxtube
or Meissner state.
Figure~\ref{fig:gap_apr_ccdk} illustrates these cases for the APR nuclear
EOS model and the CCDK model of the energy gap $\Delta$ (see below).
We see that, for $H\lesssim 10^{15}\mbox{ G}$, a large portion of the NS
interior would be in the Meissner (magnetic flux-free) state once
superconductivity sets in.
For $H\gtrsim 10^{15}\mbox{ G}$, the NS core retains its magnetic field,
either in superconducting fluxtubes or in a non-superconducting state.

The above considerations were set out in \cite{baymetal69} and explored since
then.
However, what has not been investigated quantitatively is whether magnetic flux
expulsion by the Meissner effect can occur fast enough as the NS cools soon
after formation (as $T$ drops below $\Tcp$),
although this issue is mentioned but not examined in past literature
(see, e.g., \cite{baymetal69,baym70,sinhasedrakian15}).
In order for a Meissner state to be created, magnetic field must be expelled
from the superconducting region (on the flux diffusion timescale) more
rapidly than the region grows (on the cooling timescale).
To find the former, \cite{baym70} (see also \cite{pippard50} and
V. Graber, in prep.) solves equations for flux diffusion
and energy transfer at the fixed boundary between a superconducting region
and a normal region.
This calculation yields a (modified Ohmic) diffusion timescale
(for $H/\hctwo\ll 1$; \cite{foot:hctwo})
\ba
\tohm &\approx& \tohmzero\frac{H}{2\hctwo}
 = \frac{4\pi\econd\lmag^2}{c^2}\frac{H}{2\hctwo} \nonumber\\
 &=& 4.4\times 10^{6}\mbox{ yr}\lp\frac{\econd}{10^{29}\mbox{ s$^{-1}$}}\rp
 \lp\frac{\lmag}{1\mbox{ km}}\rp^2\lp\frac{H/2\hctwo}{10^{-5}}\rp,
 \label{eq:tohm}
\ea
where $\tohmzero$ is the magnetic field diffusion/dissipation timescale
in non-superconducting matter,
$\econd$ is electrical conductivity due to scattering, and
$\lmag$ is the lengthscale over which magnetic field changes.
However, as we will show, cooling occurs much
more rapidly than flux diffusion, such that $\tohm$
may not be the correct expulsion timescale.
In superconducting matter, timescales are uncertain.
Most estimates are many orders of magnitude longer than $\tohm$
\cite{glampedakisetal11,graberetal15,graber16,dommesgusakov17,passamontietal17},
although \cite{graberetal15,graber16} derive a superconducting induction
equation with a magnetic field dissipation timescale
\be
\tsc \approx 3.9\times10^7\mbox{ yr }\lp\frac{\nnuc}{\npr}\rp^{1/6}
 \lp\frac{\lmag}{1\mbox{ km}}\rp^2 \label{eq:tsc}
\ee
that can be shorter than $\tohm$.
Note that Eq.~(\ref{eq:tsc}) uses a revised mutual friction drag
\cite{graber16}.
While the processes that lead to Eqs.~(\ref{eq:tohm}) and (\ref{eq:tsc}) may
not be the exact description for flux expulsion from superconducting matter,
$\tohm$ and $\tsc$ are the shortest known and possibly relevant timescales.
Thus each serves as a useful limiting timescale, which is sufficient for
our purposes.
We also note the important role of $\lmag$, since at small enough values,
both $\tohm$ and $\tsc$ can be very short.

In contrast, at ages $\lesssim 10^6\mbox{ yr}$, NSs cool via neutrino
emission \cite{yakovlevpethick04,potekhinetal15} over a timescale
\be
\tcool = \frac{CT}{\epsilon_\nu}
\sim 1\mbox{ yr}\lp\frac{\nn}{\nel}\rp^{1/3}\lp\frac{10^9\mbox{ K}}{T}\rp^{6},
 \label{eq:tcool}
\ee
where
$C=1.6\times 10^{20}\mbox{ erg cm$^{-3}$ K$^{-1}$ }(\nn/\nnuc)^{1/3}(T/10^9\mbox{ K})$
is neutron heat capacity,
$\epsilon_\nu\sim 3\times 10^{22}\mbox{ erg cm$^{-3}$ s$^{-1}$ }
(\nel/\nnuc)^{1/3}(T/10^9\mbox{ K})^8$ is neutrino emissivity
for modified Urca processes, and $\nn$ is neutron density.
The ratios of cooling to magnetic field diffusion timescales
$\tcool/\tohm$ and $\tcool/\tsc$ are both $\sim 10^{-8}$ for
$\lmag\approx 1\mbox{ km}$.
Clearly cooling occurs much more rapidly than flux expulsion until ages
$\gtrsim 10^6\mbox{ yr}$ when $T<10^8\mbox{ K}$.
As a result, magnetic field cannot be expelled from macroscopic regions and
is essentially frozen in nuclear matter.
A NS core remains in a (metastable) magnetized state even though the minimum
energy state is one with a flux-free configuration
(see Fig.~\ref{fig:gap_apr_ccdk}).
In the following, we describe the NS models considered here, including the
EOS and superconducting pairing gap, and present numerical results
demonstrating quantitatively the estimates given above, as well as determine
at what scale flux expulsion can occur.
We note that \cite{elfritzetal16} consider fluxtube motion from the NS core
into the crust using the formulation of \cite{graberetal15} that yields
Eq.~(\ref{eq:tsc}), whereas we consider superconductor formation/nucleation
and Meissner flux expulsion within the core at the boundary $\Tcp(\nb)$.
Finally, we emphasize that, in order to test the maximum effectiveness of flux
expulsion in comparison to cooling, we use a model which simulates slow NS
cooling and ignore effects that would lead to more rapid cooling (see below).

\section{Neutron star cooling model \label{sec:model}}

To determine the evolution of the interior temperature of an isolated NS,
we solve relativistic equations of energy balance and heat flux using
the NS cooling code described in \cite{hoetal12}.
The initial temperature is taken to be a constant $Te^\Phi=10^{10}\mbox{ K}$,
where $\Phi$ is the metric function corresponding to the gravitational
potential in the Newtonian limit \cite{shapiroteukolsky83}.
The envelope composition does not significantly affect cooling in
the core, and thus we only consider an iron composition.

We consider three nuclear EOS models that 
produce a NS with maximum mass $M>2\,\Msun$:
APR, specifically A18+$\delta v$+UIX$^\ast$ \cite{akmaletal98,pageetal04},
and BSk20 and BSk21 \cite{chameletal09,gorielyetal10,potekhinetal13}.
NS models with $M>\Mdu$ undergo the fast and efficient neutrino emission
process known as direct Urca cooling
(see, e.g., \cite{yakovlevpethick04,potekhinetal15}, for review),
and $\Mdu=1.96\,\Msun$ for APR and $\Mdu=1.59\,\Msun$ for BSk21, while
BSk20 does not produce NSs that undergo direct Urca cooling for any mass.
As we will show, modified Urca cooling is fast enough to prevent flux expulsion.
This would be even more so for NSs above the direct Urca threshold since
direct Urca cooling operates on a much faster timescale.
Thus we limit our study to $M<\Mdu$.

For proton pairing gap, we consider three models, chosen because they span
a range of densities and maximum energy gap:
AO \cite{amundsenostgaard85}, BS \cite{baldoschulze07},
and CCDK \cite{chenetal93},
and we use the gap energy parameterization from \cite{hoetal15}.
Figure~\ref{fig:gap_apr_ccdk} shows the CCDK model of the energy gap
$\Delta(\nb)$ using the APR EOS model.
The CCDK model is one that has a large maximum energy gap and spans a broad
density range.
The AO model is one that has a small maximum energy gap and smaller density
range but extends to high densities, while the BS model has a maximum energy
gap intermediate between AO and CCDK but is confined to relatively low
densities (see \cite{hoetal15}).
For the CCDK proton gap model, the criterion $\kappagl=1/\sqrt{2}$
[see Eq.~(\ref{eq:ginzburglandau})] occurs at $\nb=0.48\mbox{ fm$^{-3}$}$
for the APR EOS model and at $0.69\mbox{ fm$^{-3}$}$ and
$0.40\mbox{ fm$^{-3}$}$ for BSk20 and BSk21, respectively.

We do not consider superfluid neutrons in this work.
The dominant effect of neutron superfluidity is to enhance cooling through
neutrino emission from Cooper pairing \cite{gusakovetal04,pageetal04}.
Like the effect of direct Urca processes, this would lead to even shorter
cooling times.
It is possible that superfluid neutron-proton interactions could play a role
(see, e.g., \cite{alfordgood08,haberschmitt17}), although this probably would
not qualitatively change our conclusions.

\section{Results \label{sec:results}}
\subsection{High-mass NS with APR--CCDK models \label{sec:highm}}

To illustrate the primary findings of our work, we focus on results of one
EOS model (APR) and one superconducting proton pairing gap model (CCDK).
First we consider a high mass $1.9\,\Msun$ ($11.2\mbox{ km}$ radius) NS, in
order to probe higher densities than those of lower mass NSs.
Since $M<\Mdu$, only modified Urca and proton Cooper pairing processes
operate in the core.

\begin{figure}
\resizebox{1.0\hsize}{!}{\includegraphics{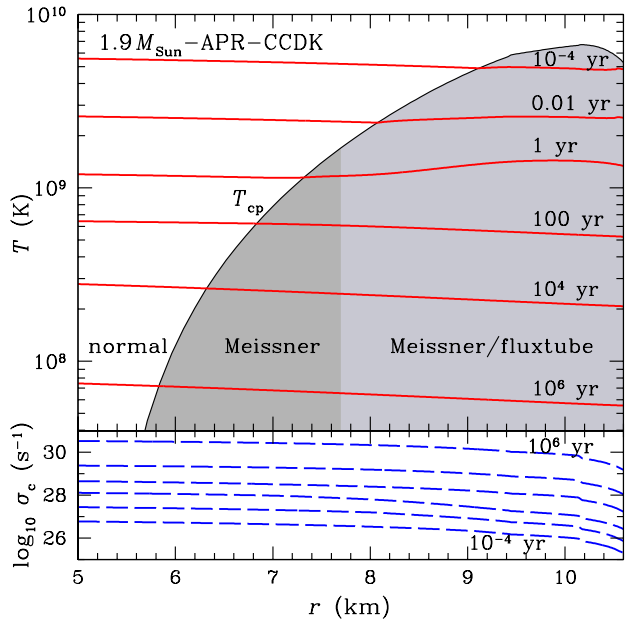}}
\caption{
Top: Temperature $T$ as a function of radius $r$.
The crust-core boundary is at $r\approx 10.7\mbox{ km}$, and total radius
is $11.2\mbox{ km}$ for this $1.9\,\Msun$ NS built using the APR EOS.
$\Tcp$ denotes the density-dependent critical temperature for onset of proton
superconductivity (when $T<\Tcp$) using the CCDK gap model.
The separation between the two shaded regions is defined by
$\kappagl=1/\sqrt{2}$ (see Fig.~\ref{fig:gap_apr_ccdk}).
Nearly horizontal curves show temperature profiles at various ages.
Bottom: Radial profile of electrical conductivity $\econd$ at
ages corresponding to temperature profiles shown in top panel.
\label{fig:cool_apr_ccdk}}
\end{figure}

Figure~\ref{fig:cool_apr_ccdk} shows the evolution of the core temperature
profile from our cooling simulation using the APR EOS and CCDK pairing gap
models.  When a NS is only several minutes old, the temperature drops below
the maximum critical temperature
[$T<\Tcp(0.2\,\mbox{ fm$^{-3}$})$; see Fig.~\ref{fig:gap_apr_ccdk}],
such that a proton superconductor begins to form at $r\sim 10\mbox{ km}$.
At subsequent times, the superconducting region grows and encompasses more
of the star.

\begin{figure}
\resizebox{1.0\hsize}{!}{\includegraphics{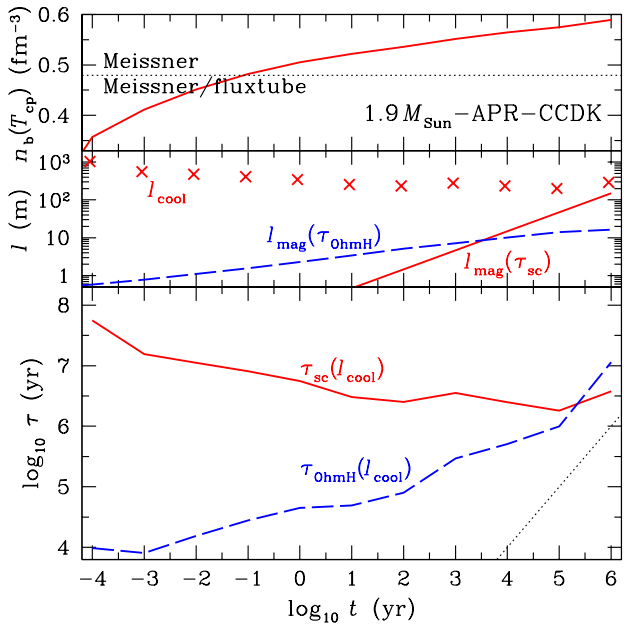}}
\caption{
Top: Density $\nb(\Tcp)$ at which onset of proton superconductivity occurs
as a function of time $t$ [since $\Tcp=T(t)$].
Horizontal dotted line (at $\nb=0.48\mbox{ fm$^{-3}$}$, where
$\kappagl=1/\sqrt{2}$) delineates regimes of Meissner and Meissner/fluxtube
states (see Fig.~\ref{fig:gap_apr_ccdk}).
Middle: Crosses are radial distance over which the superconducting region
grows at each cooling epoch and we define as cooling lengthscale $\lcool$.
Dashed line is the lengthscale $\lmag$ obtained using Eq.~(\ref{eq:tohm})
and setting $\tohm=t$, while the solid line corresponds to using
Eq.~(\ref{eq:tsc}) and setting $\tsc=t$.
Bottom: Flux diffusion timescales $\tohm$ and $\tsc$ with $\lmag=\lcool$,
where $\lcool$ is from the middle panel.
Dotted line is $\tau=t$.
In middle and bottom panels, $H/2\hctwo=10^{-5}$ is assumed.
\label{fig:scales_apr_ccdk}}
\end{figure}

The top panel of Fig.~\ref{fig:scales_apr_ccdk} shows the superconducting
boundary [at $\nb(\Tcp)$; see Fig.~\ref{fig:gap_apr_ccdk}] as a function
of time, while the middle panel shows the increase in radial extent of the
superconducting region $\lcool$ as the NS cools.
The latter is calculated at each logarithmic decade in time
($\log t_i=i$, where $i=-4,-3,\ldots,6$)
and $\lcool=r[\Tcp(t_i)]-r[\Tcp(t_{i-1})]$.
We see that the superconducting region grows by hundreds of meters every
$\Delta\log t=1$ due to cooling of the NS.
Note that we could consider shorter time intervals, so that the cooling
lengthscale is smaller, but this would necessarily imply shorter cooling
timescales as well.

In order for magnetic flux to be expelled from a Meissner region, diffusion
of magnetic field must occur over a lengthscale $\lmag$ which is greater
than the cooling lengthscale $\lcool$; otherwise magnetic flux is unable to
vacate an ever-increasing superconducting region.
We can obtain a minimum flux expulsion timescale $\tohm$
by computing the electrical conductivity $\econd$ (\cite{baymetal69b};
see bottom panel of Fig.~\ref{fig:cool_apr_ccdk}) and conservatively
setting $H/2\hctwo=10^{-5}$ (e.g., $10^{11}\mbox{ G}/10^{16}\mbox{ G}$)
and $\lmag=\lcool$ in Eq.~(\ref{eq:tohm}).
Alternatively, if we consider $\tsc$ as the flux expulsion timescale,
we find a minimum timescale by setting $\lmag=\lcool$ in Eq.~(\ref{eq:tsc}).
The bottom panel of Fig.~\ref{fig:scales_apr_ccdk} shows these timescales
$\tohm(\lcool)$ and $\tsc(\lcool)$ as functions of time $t$.
It is clear that $t\ll\tohm,\tsc$ at every epoch,
i.e., the NS cools at a much faster rate than the rate at which magnetic
flux can be expelled from superconducting regions.
Therefore magnetic field is retained within the NS core until at least
$10^6\,\mbox{yr}$.

We estimate the growing size of flux-free nucleation regions by setting
$\tohm=t$ in Eq.~(\ref{eq:tohm}) or $\tsc=t$ in Eq.~(\ref{eq:tsc}) and
solving for $\lmag(t)$.
Results are shown in the middle panel of Fig.~\ref{fig:scales_apr_ccdk}.
Flux expulsion creates Meissner state regions of size
$\sim 10\mbox{ m }$ (for $H/2\hctwo=10^{-5}$) or
$\sim 100\mbox{ m}$ after $10^6\mbox{ yr}$, depending on whether expulsion
occurs on the timescale of $\tohm$ or $\tsc$, respectively.
In addition, instead of comparing timescales, the fact that $\lcool\gg\lmag$
when $t<10^6\mbox{ yr}$ indicates the superconducting region expands by a
much larger distance than the distance over which magnetic field is expelled.

\subsection{Intermediate-mass NS with APR--CCDK models \label{sec:midm}}

\begin{figure}
\resizebox{1.0\hsize}{!}{\includegraphics{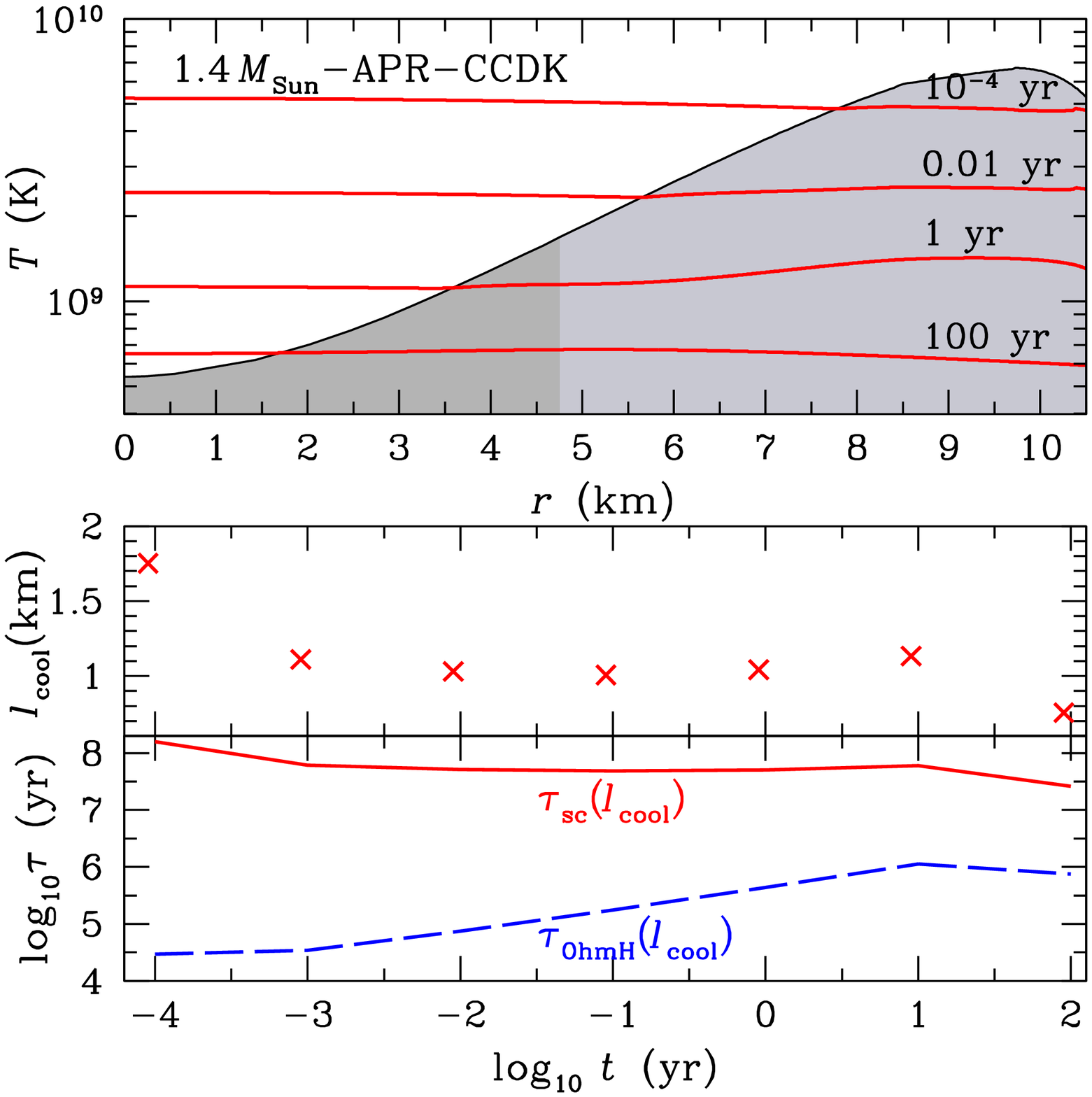}}
\caption{
Top: Temperature $T$ as a function of radius $r$.
The crust-core boundary is at $r\approx 10.6\mbox{ km}$, and total radius
is $11.6\mbox{ km}$ for this $1.4\,\Msun$ NS built using the APR EOS.
The separation between the two shaded regions is defined by
$\kappagl=1/\sqrt{2}$ (see Fig.~\ref{fig:gap_apr_ccdk}).
Nearly horizontal curves show temperature profiles at various ages.
Middle: Crosses are radial distance over which the superconducting region
grows at each cooling epoch and we define as cooling lengthscale $\lcool$.
Bottom: Flux diffusion timescales $\tohm$ and $\tsc$ with $\lmag=\lcool$,
where $\lcool$ is from the middle panel.
In middle and bottom panels, $H/2\hctwo=10^{-5}$ is assumed.
\label{fig:apr14_ccdk}}
\end{figure}

Figure~\ref{fig:apr14_ccdk} shows results for a lower mass ($1.4\,\Msun$) NS.
The central density for this NS is $\nb\approx 0.56\mbox{ fm$^{-3}$}$, which is
near the boundary defined by $\kappagl=1/\sqrt{2}$ at $0.48\mbox{ fm$^{-3}$}$.
The top panel illustrates the fact that nearly the entire core could be
in the Meissner state if magnetic flux is expelled once $T<\Tcp$
(at $t>\mbox{a few hundred years}$).
However, the middle and bottom panels show a cooling lengthscale
$\lcool\sim 1\mbox{ km}$ (larger than for a $1.9\,\Msun$ NS) and flux
diffusion timescales $\tohm,\tsc\gg t$, respectively.
Therefore magnetic flux can be expelled from the entire core only after
at least $10^7\mbox{ yr}$.

\subsection{Other EOS and proton superconducting gaps \label{sec:eossf}}

\begin{figure}
\resizebox{1.0\hsize}{!}{\includegraphics{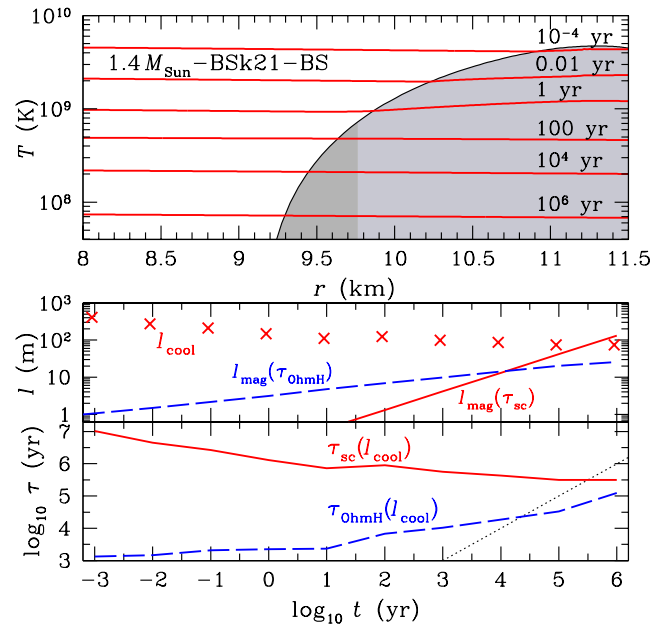}}
\caption{
Top: Temperature $T$ as a function of radius $r$.
The crust-core boundary is at $r\approx 11.5\mbox{ km}$, and total radius
is $12.6\mbox{ km}$ for this $1.4\,\Msun$ NS built using the BSk21 EOS.
The separation between the two shaded regions is defined by
$\kappagl=1/\sqrt{2}$ (see Fig.~\ref{fig:gap_apr_ccdk}).
Nearly horizontal curves show temperature profiles at various ages.
Middle: Crosses are radial distance over which the superconducting region
grows at each cooling epoch and we define as cooling lengthscale $\lcool$.
Dashed line is the lengthscale $\lmag$ obtained using Eq.~(\ref{eq:tohm})
and setting $\tohm=t$, while the solid line corresponds to using
Eq.~(\ref{eq:tsc}) and setting $\tsc=t$.
Bottom: Flux diffusion timescales $\tohm$ and $\tsc$ with $\lmag=\lcool$,
where $\lcool$ is from the middle panel.
Dotted line is $\tau=t$.
In middle and bottom panels, $H/2\hctwo=10^{-5}$ is assumed.
\label{fig:bsk21_14_bs}}
\end{figure}

We perform analogous calculations as those described above but using different
combinations of the APR, BSk20, or BSk21 nuclear EOS model and the AO, BS, or
CCDK proton pairing gap model.
The results using the CCDK model and either BSk20 or BSk21 are qualitatively
similar to those using APR.
The AO gap model is fairly broad and extends to higher densities than BS.
Results using this model are similar to those of CCDK, except times/ages at
which transitions occur later due to the lower overall $\Delta$ (and $\Tcp$).
The BS gap model is relatively narrow and centered at low densities;
for all three EOS models, the superconducting region is near the crust-core
boundary and has a radial width $\lesssim 2\mbox{ km}$.
Figure~\ref{fig:bsk21_14_bs} shows results for a $1.4\,\Msun$ NS built using
the BSk21 EOS model.
Flux expulsion from this narrow superconducting region could occur in
$\sim 10^4\,\mbox{yr}$ for relatively low ($\sim 10^{10}\mbox{ G}$)
magnetic fields.

\section{Conclusions \label{sec:discuss}}

In summary, we performed detailed cooling simulations to study the onset of
proton superconductivity in NS cores and confirmed previous estimates that
the core retains its magnetic field even though the minimum energy state
is one in which magnetic flux is expelled due to the Meissner effect.
This is because a dynamical NS cools so rapidly (even under the assumption
of slow cooling) that the superconducting region expands much faster than
the field can be expelled by any known processes.
To produce a large region in the core devoid of magnetic field, the field
must diffuse over macroscopic scales of order a kilometer or more,
and the timescale for such field diffusion is $\gtrsim 10^7\mbox{ yr}$.
At $10^6\mbox{ yr}$, the size of flux-free regions is probably $<10\mbox{ m}$
and at most $\sim 100\mbox{ m}$
(see middle panels of Figs.~\ref{fig:scales_apr_ccdk} and
\ref{fig:bsk21_14_bs}).
This suggests that there is not significant magnetic field evolution in the
core of NSs younger than at least $10^7\mbox{ yr}$
[see Eqs.~(\ref{eq:tohm}) or (\ref{eq:tsc}); see also \cite{elfritzetal16}].
Our results apply to NSs with $H>10^{11}\mbox{ G}$, including magnetars,
most of which have $H\gtrsim 10^{14}\mbox{ G}$.
Thus for observed magnetars with age $<10^5\mbox{ yr}$, there is a limit to
the amount of field decay that can occur if the magnetic field in the crust
is anchored in the core.

\begin{acknowledgments}
W.C.G.H. is grateful to K. Glampedakis, A. Schmitt, and A. Sedrakian for
discussions.  W.C.G.H. appreciates use of computer facilities at KIPAC.
W.C.G.H. and N.A. acknowledge support through grant ST/M000931/1 from STFC
in the UK.
V.G. is supported by a McGill Space Institute postdoctoral fellowship and
Trottier Chair in Astrophysics and Cosmology.
\end{acknowledgments}



\begin{thebibliography}{99}
\bibitem{foot:hctwo}
 The calculation of $\tohm$ in \cite{baym70} results in a scaling with
 respect to $H/2\hczero$, where $\hczero(T)$ is the thermodynamic critical
 field and usually satisfies $\hcone<\hczero<\hctwo$.  For simplicity, we
 instead take the scaling to be $H/2\hctwo$ since we are concerned with
 regimes where $H<\hctwo$.  Thus $\tohm$ given by Eq.~(\ref{eq:tohm})
 serves as a lower limit on the timescale.
\end{thebibliography}
\end{document}